\documentclass[prc,superscriptaddress,showpacs,floatfix]{revtex4}

\usepackage{graphicx}

\newcommand{\be}{\begin{equation}}
\newcommand{\ee}{  \end{equation}}
\newcommand{\ba}{\begin{eqnarray}}
\newcommand{\ea}{  \end{eqnarray}}

\begin{document}

\title{Abundance of Ground States with Positive Parity}

\author{T.~Papenbrock}
\affiliation{Department of Physics and Astronomy, University of Tennessee,
Knoxville, TN~37996, USA}
\affiliation{Physics Division,
Oak Ridge National Laboratory, Oak Ridge, TN 37831, USA}
\author{H.~A.~Weidenm\"uller}
\affiliation{Max-Planck Institut f\"ur Kernphysik,
D-69029 Heidelberg, Germany}
\date{\today}

\begin{abstract}
We investigate analytically and numerically a random-matrix model for
$m$ fermions occupying $\ell_1$ single-particle states with positive
parity and $\ell_2$ single-particle states with negative parity and
interacting through random two-body forces that conserve parity.  The
single-particle states are completely degenerate and carry no further
quantum numbers. We compare spectra of many-body states with positive
and with negative parity. We show that in the dilute limit defined by
$m,\ell_{1,2}\to \infty, m/\ell_{1,2}\to 0$, ground states with
positive and with negative parity occur with equal
probability. Differences in the ground-state probabilities are, thus,
a finite-size effect and are mainly due to different dimensions of the
Hilbert spaces of either parity.
\end{abstract}
\maketitle

\section{Motivation and Aim}

Johnson {et al.}~\cite{Joh98} observed that in the two-body random
ensemble (TBRE) of the nuclear shell model, ground states with spin
zero occur much more frequently than corresponds to their statistical
weight. That observation caused considerable theoretical activity (see
the reviews~\cite{Zel04} and \cite{Zhao04a}). A similar preponderance
for states with positive parity was found in
Ref.~\cite{Zha04}. We wish to explore the reason for that
preponderance. We focus attention on parity (rather than spin) because
that quantum number is analytically more easily accessible. We use a
model with spinless fermions that interact via random two-body forces.
The degenerate single-particle states carry no orbital angular
momentum quantum number but have either positive or negative
parity. The model is a modified version of EGOE(2), the embedded
two-body ensemble of Gaussian random matrices~\cite{Mon75}. We
investigate the model by using both, an analytical approach and
numerical simulations. The analytical approach evaluates traces of
powers of the Hamiltonian up to very high order and uses results of
Refs.~\cite{Pap04} and \cite{Yos06} to estimate the position of the
ground state. The numerical simulations involve diagonalization of
matrices drawn at random from the ensemble and can be done only for
Hamiltonian matrices of sufficiently small dimension whereas the
analytical approach is suited also for large-dimensional matrices.

To motivate our focus on traces of the Hamiltonian, we recall in
Section~\ref{simple} how the ground-state energy was estimated in
Refs.~\cite{Pap04} and \cite{Yos06}. That method is used and compared
with numerical simulations in Section~\ref{num}. Prior to that, we
define our model in Section~\ref{mod}. The first and second moments of
the Hamiltonian are calculated for both parities in
Section~\ref{traces}. After presenting our numerical results, we
investigate our model in the limit of large matrix dimension $N$ in
Section~\ref{dil}. We show that for $N \to \infty$, both the first and
the second moments of the Hamiltonian have the same values for either
parity. Combining that fact with the well-known result~\cite{Mon75}
that the shape of the average spectrum is asymptotically ($N \to
\infty$) Gaussian, we conclude that ground states of either parity are
equally likely. In Section~\ref{spec} we show that the strong
correlations found asymptotically for the first and second moments
extend to higher (but not to all) moments. We discuss the implications
of that result for correlations between the spectral fluctuation
properties of positive- and negative-parity states and show that the
result reinforces our conclusions. We conclude with a summary and
discussion.

\section{Simple Estimate for the Ground-state Energy}
\label{simple}

To estimate the ground-state energy, we use with proper modifications
the method introduced for states with spin in Ref.~\cite{Pap04} and
improved in Ref.~\cite{Yos06}. Let $H$ denote the Hamiltonian of the
system, ${\cal P}_{\pm}$ the projectors onto states with positive and
negative parity, respectively, and $E_{\rm ground}(\pm)$ the energies
of the lowest state with positive or negative parity, respectively. We
estimate $E_{\rm ground}(\pm)$ by writing
\be
E_{\rm ground}(\pm) = {\rm nTrace} ( H {\cal P}_{\pm} ) - r_\pm
\sigma_\pm \ .
\label{1}
\ee
The symbol ${\rm nTrace}$ stands for the normalized trace (the actual
trace divided by the dimension $N_{\pm}$ of Hilbert space), and the
width $\sigma$ is defined as
\be
\sigma^2_\pm = {\rm nTrace} ( H^2 {\cal P}_{\pm} ) \ .
\label{2}
\ee
In Ref.~\cite{Pap04} the analogue of Eq.~(\ref{1}) was used without
the first term on the right-hand side. That term was added in
Ref.~\cite{Yos06}. It represents the fluctuations of the centroid of
the spectrum. Inclusion of that term improves the agreement with
numerical simulations: the fluctuations of the parameter $r$ are
reduced. Equation~(\ref{1}) has a simple interpretation: shell-model
spectra have nearly Gaussian shape~\cite{Mon75} and are, thus,
essentially characterized by the centroid and the width. The distance
of the lowest state from the centroid of the spectrum is given by a
multiple $r_\pm$ of the width. In the case of spin, the stochastic
fluctuations of $r$ were found to be small, so that $r$ can be
considered a constant. In Ref.~\cite{Yos06}, an explicit expression
for $r$ was obtained by fitting the results of numerical
calculations. It reads
\be
r = \sqrt{0.99 \ln N + 0.36} \ .
\label{3}
\ee
We actually prefer to determine $r_\pm$ by a fit to numerical data.
In Section~\ref{num} we compare the result with Eq.~(\ref{3}). We also
compare the numerically determined probability of finding a ground
state of given parity with predictions derived from Eqs.~(\ref{1}) and
(\ref{2}).

\section{Model}
\label{mod}

We consider a system of $m$ spinless fermions distributed over a set
of degenerate single-particle states. There are $\ell_1$ states of
positive parity and $\ell_2$ states of negative parity, with
associated creation and destruction operators $a^{\dag}_{1 \mu}$,
$a^{}_{1 \mu}$ ($\mu = 1,2, \ldots, \ell_1)$ and $a^{\dag}_{2 \rho}$,
$a^{}_{2 \rho}$ ($\rho = 1,2, \ldots, \ell_2)$, respectively. The
single-particle states carry no further quantum numbers. The
many-body states of the system have positive (negative) parity if the
number $m_2$ of fermions in negative-parity states is even (odd,
respectively). The total numbers $N_+$ and $N_-$ of positive- and
negative-parity states are
\ba
N_+ = \sum_{m_1, m_2} \delta_{m_1 + m_2, m} \delta_{m_2, {\rm even}}
{\ell_1 \choose m_1 } {\ell_2 \choose m_2 } \ , \nonumber \\
N_- = \sum_{m_1, m_2} \delta_{m_1 + m_2, m} \delta_{m_2, {\rm odd}}
{\ell_1 \choose m_1 } {\ell_2 \choose m_2 } \ .
\label{4}
\ea
The Hamiltonian $H$ is a sum of two-body interactions that conserve
parity,
\ba
H &=& \frac{1}{4} \sum_{\mu \nu \rho \sigma} V^{(1)}_{\mu \nu ; \rho
\sigma} a^\dag_{1 \mu} a^\dag_{1 \nu} a^{}_{1 \sigma} a^{}_{1 \rho}
+ \frac{1}{4} \sum_{\mu \nu \rho \sigma} V^{(2)}_{\mu \nu ; \rho
\sigma} a^\dag_{2 \mu} a^\dag_{2 \nu} a^{}_{2 \sigma} a^{}_{2 \rho}
\nonumber \\
&& + \frac{1}{4} \sum_{\mu \nu \rho \sigma} X^{(1)}_{\mu \nu ; \rho
\sigma} ( a^\dag_{1 \mu} a^\dag_{1 \nu} a^{}_{2 \sigma} a^{}_{2 \rho}
+ a^\dag_{2 \rho} a^\dag_{2 \sigma} a^{}_{1 \nu} a^{}_{1 \mu} )
\nonumber \\
&& + \sum_{\mu \nu \rho \sigma} X^{(2)}_{\mu \nu; \rho \sigma}
a^{\dag}_{1 \mu} a^{\dag}_{2 \rho} a^{}_{2 \sigma} a^{}_{1 \nu} \ .
\label{5}
\ea
The ranges of the summation indices depend in an obvious way on the
creation operators and matrix elements on which they appear. The
two-body matrix elements obey the symmetry relations
\ba
V^{(1)}_{\mu \nu ;\rho \sigma} &=& V^{(1)}_{\rho \sigma; \mu \nu}
= - V^{(1)}_{\nu \mu ;\rho \sigma} = (V^{(1)}_{\mu \nu ;\rho
\sigma})^* \ , \nonumber \\ V^{(2)}_{\mu \nu ;\rho \sigma} &=&
V^{(2)}_{\rho \sigma; \mu \nu} = - V^{(2)}_{\nu \mu ;\rho \sigma} =
(V^{(2)}_{\mu \nu ;\rho \sigma})^* \ , \nonumber \\ X^{(1)}_{\mu \nu ;
\rho \sigma} &=& - X^{(1)}_{\nu \mu ; \rho \sigma} = - X^{(1)}_{\mu
\nu ; \sigma\rho} = (X^{(1)}_{\mu \nu; \rho \sigma})^* \ , \nonumber
\\ X^{(2)}_{\mu \nu; \rho \sigma} &=& (X^{(2)}_{\mu \nu; \rho
\sigma})^* \ .
\label{6}
\ea

An ensemble of Hamiltonians is obtained when we consider the matrix
elements in Eq.~(\ref{5}) as Gaussian-distributed random variables.
We assume that the $V^{(1)}_{\mu \nu ; \rho \sigma}$ are not
correlated with the $V^{(2)}_{\mu' \nu' ; \rho' \sigma'}$ and likewise
for the pairs ($V^{(i)}_{\mu \nu ; \rho \sigma}$, $X^{(k)}_{\mu' \nu'
; \rho' \sigma'}$) for $i = 1,2$ and $k = 1,2$, and for the pair
$X^{(1)}_{\mu \nu ; \rho \sigma}$, $X^{(2)}_{\mu' \nu' ; \rho'
\sigma'}$. All matrix elements have zero mean values. For the
variances, we define pairs of indices $\alpha, \beta$ by writing
$\alpha = \{ \mu \nu \}$ and likewise for $\beta$, and have for $i =
1,2$
\ba
\overline{V^{(i)}_{\alpha ; \beta} V^{(i)}_{\alpha' ; \beta'}} &=& v^2
(\delta_{\alpha \alpha'} \delta_{\beta \beta'} + \delta_{\alpha \beta'}
\delta_{\beta \alpha'}) \ , \nonumber \\
\overline{X^{(1)}_{\alpha ; \beta} X^{(1)}_{\alpha' ; \beta'}} &=& v^2
\delta_{\alpha \alpha'} \delta_{\beta \beta'} \ .
\label{7}
\ea
The bar denotes the average over the ensemble, and $\delta_{\alpha
\beta}$ stands for ($\delta_{\mu \mu'} \delta_{\nu \nu'} - \delta_{\mu
\nu'} \delta_{\nu \mu'}$), etc. The matrix elements $X^{(2)}$ do not
possess any symmetry properties and obey
\be
\overline{X^{(2)}_{\mu \nu; \rho \sigma} X^{(2)}_{\mu' \nu'; \rho'
\sigma'}} = v^2 \delta_{\mu \mu'} \delta_{\nu \nu'} \delta_{\rho \rho'}
\delta_{\sigma \sigma'} \ .
\label{7a}
\ee
Without loss of generality, we put $v^2 = 1$ in the sequel. 

\section{Calculation of ${\rm nTrace} ( H )$ and of ${\rm nTrace} (
H^2 )$.}
\label{traces}

These two traces are needed for the evaluation of Eqs.~(\ref{1}) and
(\ref{2}). The only non-vanishing contributions to the two traces
arise from terms in $H$ and in $H^2$ which leave the number of
fermions in every single-particle state unchanged. These terms are
found by using Wick contractions of the creation and annihilation
operators in the expressions for $H$ and $H^2$. We indicate the
omission of all other terms by an arrow. For $H$ we obtain
\be
H \to \frac{1}{2} \sum_{\mu \nu} V^{(1)}_{\mu \nu ; \mu \nu} n_{1 \mu}
n_{1 \nu} + \frac{1}{2} \sum_{\mu \nu} V^{(2)}_{\mu \nu ; \mu \nu}
n_{2 \mu} n_{2 \nu} + \sum_{\mu \rho} X^{(2)}_{\mu \mu; \rho \rho}
n_{1 \mu} n_{2 \rho} \ .
\label{8}
\ee
Here $n_{i \mu}$ is the number operator for state $(i \mu)$ with $i =
1,2$.

The diagonal element of $n_{1 \mu} n_{1 \nu}$ taken between one of the
states with $m_1$ fermions in positive-parity single-particle states
and $m_2$ fermions in negative-parity single-particle states
vanishes unless both states $(1 \mu)$ and $(1 \nu)$ are occupied, in
which case the matrix element equals unity. There are altogether
${\ell_1 - 2 \choose m_1 - 2} {\ell_2 \choose m_2}$ such states. We
consider separately the normalized traces over the positive-parity
and the negative-parity many-body states.  We recall that ${\cal
P}_{\pm}$ are the projection operators onto the many-body states with
positive and negative parity, respectively. We obtain
\ba
&& {\rm nTrace} ( H {\cal P}_+ ) = \frac{1}{2 N_+} \sum_{\mu \nu}
V^{(1)}_{\mu \nu ; \mu \nu} \sum_{m_1 m_2} \delta_{m_1 + m_2, m}
\delta_{m_2, {\rm even}} {\ell_1 - 2 \choose m_1 - 2} {\ell_2
\choose m_2} \nonumber \\
&& \qquad + \frac{1}{2 N_+} \sum_{\mu \nu} V^{(2)}_{\mu \nu ; \mu \nu}
\sum_{m_1 m_2} \delta_{m_1 + m_2, m} \delta_{m_2, {\rm even}} {\ell_1
\choose m_1} {\ell_2 - 2 \choose m_2 - 2} \nonumber \\
&& \qquad + \frac{1}{N_+} \sum_{\mu \rho} X^{(2)}_{\mu \mu; \rho \rho}
\sum_{m_1 m_2} \delta_{m_1 + m_2, m} \delta_{m_2, {\rm even}} {\ell_1
- 1 \choose m_1 - 1} {\ell_2 - 1 \choose m_2 - 1} \ , \nonumber \\
&& {\rm nTrace} ( H {\cal P}_- ) = \frac{1}{2 N_-} \sum_{\mu \nu}
V^{(1)}_{\mu \nu ; \mu \nu} \sum_{m_1 m_2} \delta_{m_1 + m_2, m}
\delta_{m_2, {\rm odd}} {\ell_1 - 2 \choose m_1 - 2} {\ell_2 \choose
m_2} \nonumber \\
&& \qquad + \frac{1}{2 N_-} \sum_{\mu \nu} V^{(2)}_{\mu \nu ; \mu \nu}
\sum_{m_1 m_2} \delta_{m_1 + m_2, m} \delta_{m_2, {\rm odd}} {\ell_1
\choose m_1} {\ell_2 - 2 \choose m_2 - 2} \nonumber \\
&& \qquad + \frac{1}{N_-} \sum_{\mu \rho} X^{(2)}_{\mu \mu; \rho \rho}
\sum_{m_1 m_2} \delta_{m_1 + m_2, m} \delta_{m_2, {\rm odd}} {\ell_1
- 1 \choose m_1 - 1} {\ell_2 - 1 \choose m_2 - 1} \ . \nonumber \\
\label{9}
\ea
Both traces are seen to depend on the same three uncorrelated random
variables,
\be
z_1 = \sum_{\mu \nu} V^{(1)}_{\mu \nu ; \mu \nu} \ ; \
z_2 = \sum_{\mu \nu} V^{(2)}_{\mu \nu ; \mu \nu} \ ; \
z_3 = \sum_{\mu \rho} X^{(2)}_{\mu \mu; \rho \rho} \ . 
\label{10}
\ee
As sums of uncorrelated random variables with equal Gaussian
distributions, $z_1$, $z_2$, and $z_3$ have Gaussian distributions
with mean values zero and second moments $[\ell_1 (\ell_1 - 1) / 4]$,
$[\ell_2 (\ell_2 - 1) / 4]$, and $\ell_1 \ell_2$, respectively. Thus,
the distribution of the traces in Eq.~(\ref{9}) is completely known.

The pattern that emerges in Eq.~(\ref{9}) will be seen to apply quite
generally to traces of arbitrary powers of $H$: the traces are sums of
products. The first factor in each product depends only on the random
variables and {\it is the same for both parities}. The second factor
differs for states of positive and states of negative parity but {\it
is independent of the random variables}. That general pattern will be
decisive for our understanding of the preponderance of ground states
with positive parity.

We turn to ${\rm Trace} \ ( H^2 )$. The following terms yield non-zero
contributions: the square of the first term on the right-hand side of
Eq.~(\ref{5}), the square of the second term, the product of the first
and the second term, the square of the third term, and the square of
the fourth term. We consider these terms in turn.

In the square of the first term, there appear the two matrix elements
$V^{(1)}$ with their associated creation and annihilation operators.
Wick contraction is possible in three different ways: (i) We contract
the two creation and the two annihilation operators associated with
the same matrix element. That is the same procedure as used in
formula~(\ref{8}) and yields a total of 4 contraction patterns. (ii)
We contract one of the two creation operators associated with the
first matrix element with an annihilation operator associated with the
same matrix element, and the other with an annihilation operator
associated with the second matrix element. That yields a total of 16
contraction patterns. (iii) We contract the two creation operators
associated with the first matrix element with the two annihilation
operators associated with the second matrix element. That yields a
total of 4 contraction patterns. It is straightforward to check that
because of the fermionic anticommutation rules and the symmetry
properties~(\ref{6}), the different contraction patterns in each of
the three groups yield identical results. For the square of the second
term on the right-hand side of Eq.~(\ref{5}), these considerations
apply likewise. For the product of the first and second term, only
the contraction patterns used in formula~(\ref{8}) are possible. In
the square of the third term on the right-hand side of Eq.~(\ref{5}),
only the product of the two terms in round brackets gives a
non-vanishing contribution, with obvious contraction patterns. In the
square of the fourth term, there are the same three possibilities as
in the square of the first term. Altogether this yields
\ba
H^2 &\to& \bigg( \sum_{i = 1}^2 \frac{1}{2} V^{(i)}_{\alpha \beta ;
\alpha \beta} n_{i \alpha} n_{i \beta} \bigg)^2 + \sum_{i = 1}^2
\sum_{\alpha \beta \beta'} V^{(i)}_{\alpha \beta ; \alpha \beta}
V^{(i)}_{\alpha \beta' ; \alpha \beta'} n_{i \alpha} n_{i \beta}
n_{i \beta'} \nonumber \\
&& + \sum_{i = 1}^2 \sum_{\alpha \beta \alpha' \beta'} V^{(i)}_{\alpha
\beta ; \alpha' \beta} V^{(i)}_{\alpha \beta' ; \alpha' \beta'} n_{i
\alpha} n_{i \beta} ( 1 - n_{i \alpha'} ) n_{i \beta'} \nonumber \\   
&& + \sum_{i = 1}^2 \bigg( \frac{1}{2} \sum_{\alpha \beta}
(V^{(i)}_{\alpha \beta ; \alpha \beta})^2 n_{i \alpha} n_{i \beta}
+ \sum_{\alpha \beta \alpha'} ( V^{(i)}_{\alpha \beta ; \alpha'
\beta})^2 n_{i \alpha} n_{i \beta} ( 1 - n_{i \alpha'}) \nonumber \\
&& \qquad + \frac{1}{4} \sum_{\alpha \beta \alpha' \beta'}
( V^{(i)}_{\alpha \beta ; \alpha' \beta'})^2 n_{i \alpha} n_{i \beta}
( 1 - n_{i \alpha'} ) ( 1 - n_{i \beta'} ) \bigg) \nonumber \\
&& + \frac{1}{4} \sum_{\alpha \beta \alpha' \beta'} (X^{(1)}_{\alpha
\beta ; \alpha' \beta'})^2 \bigg( n_{1 \alpha} n_{1 \beta} ( 1 - n_{2
\alpha'} ) ( 1 - n_{2 \beta'} ) \nonumber \\
&& \qquad + ( 1 - n_{1 \alpha} ) ( 1 - n_{1 \beta} ) n_{2 \alpha'}
n_{2 \beta'} \bigg) \nonumber \\
&& + \bigg( \sum_{\mu \rho} X^{(2)}_{\mu \mu; \rho \rho} n_{1 \mu}
n_{2 \rho} \bigg)^2 \nonumber \\
&& + \sum_{\mu \nu \rho \sigma} X^{(2)}_{\mu \nu; \rho \rho}
X^{(2)}_{\nu \mu; \sigma \sigma} n_{1 \mu} (1 - n_{1 \nu}) n_{2 \rho}
n_{2 \sigma} \nonumber \\
&& + \sum_{\mu \nu \rho \sigma} X^{(2)}_{\mu \mu; \rho \sigma}
X^{(2)}_{\nu \nu; \sigma \rho} n_{1 \mu} n_{1 \nu} n_{2 \rho} (1 -
n_{2 \sigma}) \nonumber \\
&& + \sum_{\mu \nu \rho \sigma} X^{(2)}_{\mu \nu; \rho \sigma}
X^{(2)}_{\nu \mu; \sigma \rho} n_{1 \mu} (1 - n_{1 \nu}) n_{2 \rho}
(1 - n_{2 \sigma}) \ .
\label{11}
\ea
Before working out the trace of this expression, it is useful to
rearrange it in such a way that in all summations no two summation
indices take the same values. This yields
\ba
H^2 &\to& \sum_{i = 1}^2 \bigg( 2 \sum_{\alpha \beta} ( V^{(i)}_{
\alpha \beta ; \alpha \beta} )^2 n_{i \alpha} n_{i \beta} \nonumber \\
&& + 2 \sum_{\alpha \beta \beta'} ( V^{(i)}_{\alpha \beta ; \beta'
\beta} )^2 n_{i \alpha} n_{i \beta} ( 1 - n_{i \beta'} ) \nonumber \\
&& + 2 \sum_{\alpha \beta \beta'} ( 1 - \delta_{\beta \beta'} )
V^{(i)}_{\alpha \beta ; \alpha \beta} V^{(i)}_{\alpha \beta' ; \alpha
\beta'} n_{i \alpha} n_{i \beta} n_{i \beta'} \nonumber \\
&& + \frac{1}{4} \sum_{\alpha \beta \alpha' \beta'} ( V^{(i)}_{
\alpha \beta ; \alpha' \beta'} )^2 n_{i \alpha} n_{i \beta} ( 1 -
n_{i \alpha'} ) ( 1 - n_{i \beta'} ) \nonumber \\
&& + \frac{1}{4} \sum_{\alpha \beta \alpha' \beta'} ( 1 -
\delta_{\alpha \alpha'} ) ( 1 - \delta_{\alpha \beta'} ) ( 1 -
\delta_{\beta \alpha'} ) ( 1 - \delta_{\beta \beta'} ) \nonumber \\
&& \qquad \times V^{(i)}_{\alpha \beta ; \alpha \beta} V^{(i)}_{
\alpha' \beta' ; \alpha' \beta'} n_{i \alpha} n_{i \beta} n_{i \alpha'}
n_{i \beta'} \nonumber \\
&& + \sum_{\alpha \beta \alpha' \beta'} ( 1 - \delta_{\alpha \alpha'} )
( 1 - \delta_{\beta \beta'} ) V^{(i)}_{\alpha \beta ; \alpha' \beta}
V^{(i)}_{\alpha \beta' ; \alpha' \beta'} \nonumber \\
&& \qquad \times n_{i \alpha} n_{i \beta} ( 1 - n_{i \alpha'} ) n_{i
\beta'} \bigg) \nonumber \\
&& + \frac{1}{2} \sum_{\alpha \beta \alpha' \beta'} V^{(1)}_{\alpha
\beta ; \alpha \beta} V^{(2)}_{\alpha' \beta' ; \alpha' \beta'} n_{1
\alpha} n_{1 \beta} n_{2 \alpha'} n_{2 \beta'} \nonumber \\
&& + \frac{1}{4} \sum_{\alpha \beta \alpha' \beta'} (X^{(1)}_{\alpha
\beta; \alpha' \beta'})^2 \bigg( n_{1 \alpha} n_{1 \beta} ( 1 -
n_{2 \alpha'} ) ( 1 - n_{2 \beta'} ) \nonumber \\
&& \qquad + ( 1 - n_{1 \alpha} ) ( 1 - n_{1 \beta} ) n_{2 \alpha'}
n_{2 \beta'} \bigg) \nonumber \\
&& + \sum_{\mu \rho} (X^{(2)}_{\mu \mu; \rho \rho})^2 n_{1 \mu} n_{2
\rho} \nonumber \\
&& + \sum_{\mu \mu' \rho} (1 - \delta_{\mu \mu'}) X^{(2)}_{\mu \mu;
\rho \rho} X^{(2)}_{\mu' \mu'; \rho \rho} n_{1 \mu} n_{1 \mu'} n_{2
\rho} \nonumber \\
&& + \sum_{\mu \rho \rho'} (1 - \delta_{\rho \rho'}) X^{(2)}_{\mu
\mu; \rho \rho} X^{(2)}_{\mu \mu; \rho' \rho'} n_{1 \mu} n_{2 \rho}
n_{2 \rho'} \nonumber \\
&& + \sum_{\mu \mu' \rho \rho'} (1 - \delta_{\mu \mu'}) (1 -
\delta_{\rho \rho'}) X^{(2)}_{\mu \mu; \rho \rho} X^{(2)}_{\mu' \mu';
\rho' \rho'} n_{1 \mu} n_{1 \mu'} n_{2 \rho} n_{2 \rho'} \nonumber \\
&& + \sum_{\mu \nu \rho} X^{(2)}_{\mu \nu; \rho \rho} X^{(2)}_{\nu
\mu; \rho \rho} n_{1 \mu} (1 - n_{1 \nu}) n_{2 \rho} \nonumber \\
&& + \sum_{\mu \nu \rho \sigma} (1 - \delta_{\rho \sigma})
X^{(2)}_{\mu \nu; \rho \rho} X^{(2)}_{\nu \mu; \sigma \sigma} n_{1
\mu} (1 - n_{1 \nu}) n_{2 \rho} n_{2 \sigma} \nonumber \\
&& + \sum_{\mu \rho \sigma} X^{(2)}_{\mu \mu; \rho \sigma}
X^{(2)}_{\mu \mu; \sigma \rho} n_{1 \mu} n_{2 \rho} (1 - n_{2 \sigma})
\nonumber \\
&& + \sum_{\mu \nu \rho \sigma} (1 - \delta_{\mu \nu}) X^{(2)}_{\mu
\mu; \rho \sigma} X^{(2)}_{\nu \nu; \sigma \rho} n_{1 \mu} n_{1 \nu}
n_{2 \rho} (1 - n_{2 \sigma}) \nonumber \\
&& + \sum_{\mu \nu \rho \sigma} X^{(2)}_{\mu \nu; \rho \sigma}
X^{(2)}_{\nu \mu; \sigma \rho} n_{1 \mu} (1 - n_{1 \nu}) n_{2 \rho}
(1 - n_{2 \sigma}) \ .
\label{12}
\ea
In calculating the trace, we observe that the number of nonequal
summation indices in the terms on the right-hand side of
Eq.~(\ref{12}) determines the weight factors. The result
is
\ba
{\rm nTrace} ( H^2 {\cal P}_+ ) &=& \sum_{m_1 m_2} \delta_{m_1
+ m_2, m} \delta_{m_2, {\rm even}} \nonumber \\
&& \times \bigg\{ \frac{2}{N_+} \sum_{\alpha \beta} ( V^{(1)}_{\alpha
\beta ; \alpha \beta} )^2 {\ell_1 - 2 \choose m_1 - 2} {\ell_2 \choose
m_2} \nonumber \\
&& + \frac{2}{N_+} \sum_{\alpha \beta} ( V^{(2)}_{\alpha \beta ;
\alpha \beta} )^2 {\ell_1 \choose m_1} {\ell_2 - 2 \choose m_2 - 2}
\nonumber \\
&& + \frac{2}{N_+} \sum_{\alpha \beta \beta'} ( 1 - \delta_{\alpha
\beta'} ) ( V^{(1)}_{\alpha \beta ; \beta' \beta} )^2 {\ell_1 - 3
\choose m_1 - 2} {\ell_2 \choose m_2} \nonumber \\
&& + \frac{2}{N_+} \sum_{\alpha \beta \beta'} ( 1 - \delta_{\alpha
\beta'} ) ( V^{(2)}_{\alpha \beta ; \beta' \beta} )^2 {\ell_1 \choose
m_1} {\ell_2 - 3 \choose m_2 - 2} \nonumber \\
&& + \frac{2}{N_+} \sum_{\alpha \beta \beta'} ( 1 - \delta_{\beta
\beta'} ) V^{(1)}_{\alpha \beta ; \alpha \beta} V^{(1)}_{\alpha \beta'
; \alpha \beta'} \nonumber \\
&& \qquad \times {\ell_1 - 3 \choose m_1 - 3} {\ell_2 \choose m_2}
\nonumber \\
&& + \frac{2}{N_+} \sum_{\alpha \beta \beta'} ( 1 - \delta_{\beta
\beta'} ) V^{(2)}_{\alpha \beta ; \alpha \beta} V^{(2)}_{\alpha \beta'
; \alpha \beta'} \nonumber \\
&& \qquad \times {\ell_1 \choose m_1} {\ell_2 - 3 \choose m_2 - 3}
\nonumber \\
&& + \frac{1}{4 N_+} \sum_{\alpha \beta \alpha' \beta'} (1 -
\delta_{\alpha \alpha'} ) (1 - \delta_{\alpha \beta'} ) (1 -
\delta_{\beta \alpha'} ) (1 - \delta_{\beta \beta'} ) \nonumber \\
&& \qquad \times ( V^{(1)}_{\alpha \beta ; \alpha' \beta'} )^2
{\ell_1 - 4 \choose m_1 - 2} {\ell_2 \choose m_2} \nonumber \\
&& + \frac{1}{4 N_+} \sum_{\alpha \beta \alpha' \beta'} (1 -
\delta_{\alpha \alpha'} ) (1 - \delta_{\alpha \beta'} ) (1 -
\delta_{\beta \alpha'} ) (1 - \delta_{\beta \beta'} ) \nonumber \\
&& \qquad \times ( V^{(2)}_{\alpha \beta ; \alpha' \beta'} )^2
{\ell_1 \choose m_1} {\ell_2 - 4 \choose m_2 - 2} \nonumber \\
&& + \frac{1}{4 N_+} \sum_{\alpha \beta \alpha' \beta'} ( 1 -
\delta_{\alpha \alpha'} ) ( 1 - \delta_{\alpha \beta'} ) ( 1 -
\delta_{\beta \alpha'} ) ( 1 - \delta_{\beta \beta'} ) \nonumber \\
&& \qquad \times V^{(1)}_{\alpha \beta ; \alpha \beta} V^{(1)}_{
\alpha' \beta' ; \alpha' \beta'} {\ell_1 - 4 \choose m_1 - 4}
{\ell_2 \choose m_2} \nonumber \\
&& + \frac{1}{4 N_+} \sum_{\alpha \beta \alpha' \beta'} ( 1 -
\delta_{\alpha \alpha'} ) ( 1 - \delta_{\alpha \beta'} ) ( 1 -
\delta_{\beta \alpha'} ) ( 1 - \delta_{\beta \beta'} ) \nonumber \\
&& \qquad \times V^{(2)}_{\alpha \beta ; \alpha \beta} V^{(2)}_{
\alpha' \beta' ; \alpha' \beta'} {\ell_1 \choose m_1} {\ell_2 - 4
\choose m_2 - 4} \nonumber \\
&& + \frac{1}{N_+} \sum_{\alpha \beta \alpha' \beta'} ( 1 -
\delta_{\alpha \alpha'} ) ( 1 - \delta_{\beta \beta'} )
V^{(1)}_{\alpha \beta ; \alpha' \beta} V^{(1)}_{\alpha \beta' ;
\alpha' \beta'} \nonumber \\
&& \qquad \times {\ell_1 - 4 \choose m_1 - 3} {\ell_2 \choose m_2}
\nonumber \\
&& + \frac{1}{N_+} \sum_{\alpha \beta \alpha' \beta'} ( 1 -
\delta_{\alpha \alpha'} ) ( 1 - \delta_{\beta \beta'} )
V^{(2)}_{\alpha \beta ; \alpha' \beta} V^{(2)}_{\alpha \beta' ;
\alpha' \beta'} \nonumber \\
&& \qquad \times {\ell_1 \choose m_1} {\ell_2 - 4 \choose m_2 - 3}
\nonumber \\
&& + \frac{1}{2 N_+} \sum_{\alpha \beta \alpha' \beta'}
V^{(1)}_{\alpha \beta ; \alpha \beta} V^{(2)}_{\alpha' \beta' ;
\alpha' \beta'} {\ell_1 - 2 \choose m_1 - 2} {\ell_2 - 2 \choose
m_2 - 2} \nonumber \\
&& + \frac{1}{4 N_+} \sum_{\alpha \beta \alpha' \beta'}
(X^{(1)}_{\alpha \beta ; \alpha' \beta'})^2 \bigg[ {\ell_1 - 2
\choose m_1 - 2} {\ell_2 - 2 \choose m_2} \nonumber \\
&& \qquad + {\ell_1 - 2 \choose m_1} {\ell_2 - 2 \choose m_2 - 2}
\bigg] \nonumber \\
&& + \frac{1}{N_+} \sum_{\mu \rho} (X^{(2)}_{\mu \mu; \rho \rho})^2
{\ell_1 - 1 \choose m_1 - 1} {\ell_2 - 1 \choose m_2 - 1}
\nonumber \\
&& + \frac{1}{N_+} \sum_{\mu \mu' \rho} (1 - \delta_{\mu \mu'})
X^{(2)}_{\mu \mu; \rho \rho} X^{(2)}_{\mu' \mu'; \rho \rho} {\ell_1
- 2 \choose m_1 - 2} {\ell_2 - 1 \choose m_2 - 1} \nonumber \\
&& + \frac{1}{N_+} \sum_{\mu \rho \rho'} (1 - \delta_{\rho \rho'})
X^{(2)}_{\mu \mu; \rho \rho} X^{(2)}_{\mu \mu; \rho' \rho'} {\ell_1
- 1 \choose m_1 - 1} {\ell_2 - 2 \choose m_2 - 2} \nonumber \\
&& + \frac{1}{N_+} \sum_{\mu \mu' \rho \rho'} (1 - \delta_{\mu
\mu'}) (1 - \delta_{\rho \rho'}) X^{(2)}_{\mu \mu; \rho \rho}
X^{(2)}_{\mu' \mu'; \rho' \rho'} \nonumber \\
&& \qquad \times {\ell_1 - 2 \choose m_1 - 2} {\ell_2 - 2 \choose
m_2 - 2} \nonumber \\
&& + \frac{1}{N_+} \sum_{\mu \nu \rho} (1 - \delta_{\mu \nu})
X^{(2)}_{\mu \nu; \rho \rho} X^{(2)}_{\nu \mu; \rho \rho} {\ell_1 -
2 \choose m_1 - 1} { \ell_2 - 1 \choose m_2 - 1} \nonumber \\
&& + \frac{1}{N_+} \sum_{\mu \nu \rho \sigma} (1 - \delta_{\mu \nu})
(1 - \delta_{\rho \sigma}) X^{(2)}_{\mu \nu; \rho \rho} X^{(2)}_{\nu
\mu; \sigma \sigma} \nonumber \\
&& \qquad \times { \ell_1 - 2 \choose m_1 - 1} { \ell_2 - 2 \choose
m_2 - 2} \nonumber \\
&& + \frac{1}{N_+} \sum_{\mu \rho \sigma} (1 - \delta_{\rho \sigma})
X^{(2)}_{\mu \mu; \rho \sigma} X^{(2)}_{\mu \mu; \sigma \rho}
{\ell_1 - 1 \choose m_1 - 1} { \ell_2 - 2 \choose m_2 - 1}
\nonumber \\
&& + \frac{1}{N_+} \sum_{\mu \nu \rho \sigma} (1 - \delta_{\mu \nu})
(1 - \delta_{\rho \sigma}) X^{(2)}_{\mu \mu; \rho \sigma}
X^{(2)}_{\nu \nu; \sigma \rho} \nonumber \\
&& \qquad \times { \ell_1 - 2 \choose m_1 - 2} { \ell_2 - 2 \choose
m_2 - 1} \nonumber \\
&& + \frac{1}{N_+} \sum_{\mu \nu \rho \sigma} (1 - \delta_{\mu \nu})
(1 - \delta_{\rho \sigma}) X^{(2)}_{\mu \nu; \rho \sigma}
X^{(2)}_{\nu \mu; \sigma \rho} \nonumber \\
&& \qquad \times { \ell_1 - 2 \choose m_1 - 1 } { \ell_2 - 2 \choose
m_2 - 1 } \bigg\} \ .
\label{13}
\ea
For ${\rm nTrace} ( H^2 {\cal P}_- )$ we find exactly the same
expression except that the second Kronecker delta in the first line on
the right-hand side of Eq.~(\ref{13}) is replaced by $\delta_{m_2,
{\rm odd}}$, and that $N_+$ is replaced everywhere by $N_-$.

As in the case of ${\rm Trace} H$, the trace of $H^2$ is a sum of
terms each of which is the product of two factors. One factor depends
only on the random variables and is the same for both parities. The
distribution of these factors can be worked out. That is not done
here. Some of the factors are correlated with each other. The other
factor is a weight factor that is a sum over products of binomial
factors. It does not depend on the random variables and is not
obviously the same for the two parities. Our result would not apply in
the case of states with spin where the linear or bilinear forms
containing the random variables will depend on the total spin.
Assuming that Eqs.~(\ref{1}) to (\ref{3}) hold, we conclude that a
preponderance of ground states with even parity - if it exists - can
have only one of two causes: it may be due to differences between the
non-statistical weight factors, or to differences in the scale
factors $r_+$ and $r_-$. (We recall that according to Eq.~(\ref{3}),
the latter depend on the matrix dimensions $N_{\pm}$.)

\section{Numerical Results}
\label{num}

\begin{figure}[h]
\includegraphics[width=0.45\textwidth,clip=]{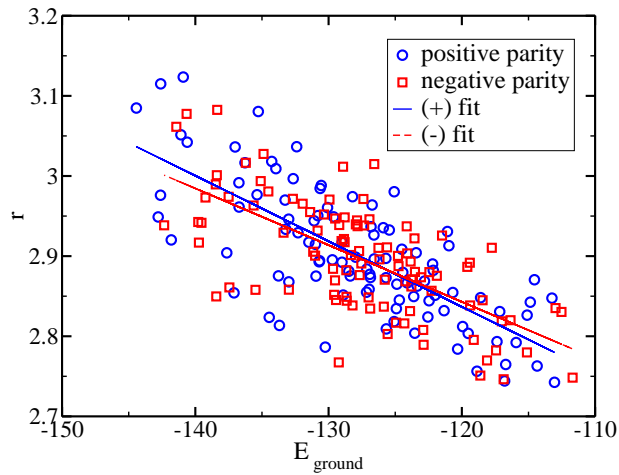}
\caption{(Color online) Scale factors $r_\pm$ of Eq.~(\ref{1}) for a
system of $m=9$ fermions on $\ell_1=9$ single-particle orbitals with
positive parity and $\ell_2=9$ orbitals with negative parity as a
function of the ground-state energies $E_{\rm ground}(\pm)$.}
{}\label{fig1}
\end{figure}

For a test of Eq.(\ref{1}), we perform numerical simulations. To this
purpose, we consider several systems that differ in the parameters
$\ell_1$, $\ell_2$, and $m$.  For each set of parameters, we set up the
matrix corresponding to the Hamiltonian~(\ref{5}) in a space of Slater
determinants. The Gaussian-distributed two--body matrix elements are
computed by a pseudo-random number generator, and the ground-state
energies $E_{\rm ground}(\pm)$ are obtained from a numerical
diagonalization of the Hamiltonian matrix. For the largest-dimensional
matrices, we employ the {\sc Arpack} package~\cite{arpack} in the
diagonalization. In addition to the ground-state energy we also
compute the normalized traces ${\rm nTrace}(H^k{\cal P}_\pm)$ for
$k=1,2$.  Our ensemble consists of 100 random Hamiltonians for each
set of parameters $\ell_1$, $\ell_2$, and $m$, and we record the
ground-state energies $E_{\rm ground}(\pm)$ and the first two moments
${\rm nTrace}(H^k{\cal P}_\pm)$ (with $k=1,2$) of the parity-projected
Hamiltonian for each member of the ensemble. We employ Eq.~(\ref{1})
and determine the scale factors $r_\pm$ that relate the ground-state
energy to the first and second moment of the Hamiltonian by fit. An
example is shown for the set of parameters $m=\ell_1=\ell_2=9$ in
Fig.~\ref{fig1}. The results obtained for the scale factors (with
their rms variances) are shown in Table~\ref{tab1}.  The table also
shows the probability $p_+$ that the ground state has positive
parity. Inspection of Table~\ref{tab1} shows that the parity of the
ground state is very sensitive to $r_\pm$. A small difference in the
scale factors $r_\pm$ is more strongly correlated with the parity of
the ground state than a small difference in the numbers $N_\pm$ of
many-body basis states.

Once the scale factors are determined, we can test how well the
right-hand side of Eq.~(\ref{1}) can be used to determine the parity
of the ground state. Our results show that the application of
Eq.~(\ref{1}) with an energy-independent scale factor does not yield
reliable predictions. Indeed, Fig.~\ref{fig1} suggests that a linear
relation $r_\pm(E_{\rm ground}(\pm)) = a_\pm +b_\pm E_{\rm
ground}(\pm)$ should describe the data more accurately.  Again, we
determine the coefficients $a_\pm$ and $b_\pm$ by fit, and then employ
the right-hand side of Eq.~(\ref{1}) with the energy-dependent scale
factor to determine the parity of the ground state as
\be
{\rm sign}\left(
{{\rm nTr} H{\cal P}_- -a_- \sigma_-\over 1+ b_-\sigma_-} -
{{\rm nTr} H{\cal P}_+ -a_+ \sigma_+\over 1+ b_+\sigma_+}
\right) \ .
\ee
Though this estimate is not correct for each individual member of the
ensemble, it yields reasonably reliable predictions for the estimated
probability $p_+({\rm est})$ of finding a ground-state with positive
parity.  Our results for this probability are shown in the last column
of Table~\ref{tab1}.

\begin{table}[h]
\begin{tabular}{@{}|r|r|r|r|r|c|c|c|c|c|c|@{}}\hline
$\ell_1$ & $\ell_2$ & $m$ & $N_+$ & $N_-$ & $r_+$ & $r_-$ & $p_+$ & $p_+$({\rm est}) \\
\hline
 6 &  6 & 6 &   452 &   472 & 2.39 $\pm$ 0.12 & 2.43 $\pm$ 0.12 & 0.18 & 0.00 \\
 7 &  7 & 5 &  1001 &  1001 & 2.42 $\pm$ 0.08 & 2.42 $\pm$ 0.08 & 0.47 & 0.49 \\
 7 &  7 & 6 &  1484 &  1519 & 2.51 $\pm$ 0.08 & 2.55 $\pm$ 0.08 & 0.20 & 0.03 \\
 7 &  7 & 7 &  1716 &  1716 & 2.61 $\pm$ 0.08 & 2.60 $\pm$ 0.08 & 0.48 & 0.56 \\
 9 &  9 & 5 &  4284 &  4284 & 2.47 $\pm$ 0.06 & 2.47 $\pm$ 0.06 & 0.55 & 0.54 \\
10 &  8 & 5 &  4312 &  4256 & 2.48 $\pm$ 0.06 & 2.46 $\pm$ 0.05 & 0.84 & 1.00 \\
 9 &  7 & 8 &  6435 &  6435 & 2.77 $\pm$ 0.07 & 2.76 $\pm$ 0.08 & 0.54 & 0.58 \\
 8 &  8 & 8 &  6470 &  6400 & 2.78 $\pm$ 0.07 & 2.74 $\pm$ 0.08 & 0.83 & 1.00 \\
10 &  6 & 8 &  6390 &  6480 & 2.73 $\pm$ 0.08 & 2.77 $\pm$ 0.09 & 0.18 & 0.00 \\
 9 &  9 & 9 & 24310 & 24310 & 2.90 $\pm$ 0.08 & 2.90 $\pm$ 0.07 & 0.52 & 0.57 \\
 8 & 10 & 9 & 24240 & 24380 & 2.87 $\pm$ 0.07 & 2.91 $\pm$ 0.07 & 0.20 & 0.00 \\
 7 & 11 & 9 & 24310 & 24310 & 2.88 $\pm$ 0.07 & 2.89 $\pm$ 0.07 & 0.50 & 0.27 \\
\hline
\end{tabular}
\caption{Results of numerical simulations.  Here, $m$, $\ell_1$, and
$\ell_2$ denote the number of fermions and the number of
single-particle levels with positive and negative parity,
respectively. $N_\pm$ is the number of many-body states with the
indicated parity, and $r_\pm$ denote the scale factors. $p_+$ denotes
the probability that the ground state has positive parity, while
$p_+({\rm est })$ is the probability that the estimated ground state
has positive parity (based on Eq.~(\ref{1}) with
a scale factor that is a polynomial of degree one in the energy).}
\label{tab1}
\end{table}

\section{Dilute Limit}
\label{dil}

In canonical random-matrix theory, attention is usually focused on the
limit of large matrix dimension. We follow suit by considering our
model in the ``dilute limit''~\cite{Mon75} defined by
$\ell_{1,2},m\to\infty$ and $m/\ell_{1,2}\to 0$. In practice, we
compute the leading order of expressions of interest under the strong
conditions $1\ll m\ll\ell_{1,2}$. We show that the weight factors
appearing in the traces of $H^k$ with $k = 1, 2$ for positive and for
negative parity become asymptotically equal. That statement holds not
only for $\ell_1 = \ell_2$ but also for $\ell_1 \neq \ell_2$.

Equations~(\ref{9}) and (\ref{13}) show that for the positive parity
states, all weight factors have the general form
\be
\sum_{m_1, m_2} \delta_{m_1 + m_2, m} \delta_{m_2, {\rm even}} \frac{1}{N_+}
{\ell_1 - \alpha_1 \choose m_1 - \beta_1} {\ell_2 - \alpha_2 \choose
m_2 - \beta_2} \ ,
\label{33}
\ee
with $\alpha_1, \alpha_2, \beta_1, \beta_2$ small positive integers.
We evaluate the sums in Eq.~(\ref{33}) and the corresponding sums
defining $N_+$ with the help of Stirling's formula, $n! \approx \exp
\{ n \ln n - n \}$. With $\mu$ integer, we write $m_2 = 2 \mu, m_1 = m
- 2 \mu$ and have for the numerator of Eq.~(\ref{33}) (all terms
except for $(N_+)^{-1}$)
\ba
&& \sum_\mu \exp \bigg\{ (\ell_1 - \alpha_1) \ln (\ell_1 - \alpha_1)
- (m - 2 \mu - \beta_1) \ln (m - 2 \mu - \beta_1) \nonumber \\
&& \qquad - (\ell_1 - \alpha_1 - m + 2 \mu + \beta_1) \ln (\ell_1 -
\alpha_1 - m + 2 \mu + \beta_1) \nonumber \\
&& \qquad + (\ell_2 - \alpha_2) \ln (\ell_2 - \alpha_2) - (2 \mu -
\beta_2) \ln (2 \mu - \beta_2) \nonumber \\
&& \qquad - (\ell_2 - \alpha_2 - 2 \mu + \beta_2) \ln (\ell_2 -
\alpha_2 - 2 \mu + \beta_2) \bigg\} \ .
\label{34}
\ea
We write the sum as an integral over $\mu$. The integrand assumes its
maximum value at
\be
\mu^{(0)}_+ = \frac{1}{2} \ \frac{(m - \beta_1)(\ell_2 - \alpha_2) +
\beta_2 (\ell_1 - \alpha_1)}{\ell_1 - \alpha_1 + \ell_2 - \alpha_2} \ .
\label{35}
\ee
With $\delta \mu = \mu - \mu_0$, expansion around the maximum yields
the negative-definite quadratic form
\ba
&& - \frac{2 (\delta \mu)^2}{m - 2 \mu^{(0)}_+ - \beta_1}
- \frac{2 (\delta \mu)^2}{\ell_1 - \alpha_1 - m + 2 \mu^{(0)}_+ +
\beta_1} - \frac{2 (\delta \mu)^2}{2 \mu^{(0)}_+ - \beta_2}
\nonumber \\
&& - \frac{2 (\delta \mu)^2}{\ell_2 - \alpha_2 - 2 \mu^{(0)}_+ +
\beta_2} = \frac{1}{2} \frac{(\delta \mu)^2}{\tau^2} \ .
\label{36}
\ea
Here the last equation defines the width $\tau$. For $\ell_1 \gg 1,
\ell_2 \gg 1, m \gg 1$ we have $\mu_0 \gg 1$. For the dilute limit, we 
neglect terms of higher 
order, and the resulting integral is Gaussian. We extend the
integration from $- \infty$ to $+ \infty$. The numerator of
expression~(\ref{33}) becomes
\ba
&& \sqrt{2 \pi} \tau \exp \bigg\{ (\ell_1 - \alpha_1) \ln (\ell_1 -
\alpha_1) + (\ell_2 - \alpha_2) \ln (\ell_2 - \alpha_2) \bigg\}
\nonumber \\
&& \times \exp \bigg\{ - (m - 2 \mu^{(0)}_+ - \beta_1) \ln (m - 2
\mu^{(0)}_+ - \beta_1) \bigg\} \nonumber \\
&& \times \exp \bigg\{ - (2 \mu^{(0)}_+ - \beta_2) \ln (2 \mu^{(0)}_+
- \beta_2) \bigg\} \nonumber \\ &&
\times \exp \bigg\{ - (\ell_1 - \alpha_1 - m + 2 \mu^{(0)}_+ +
\beta_1) \ln (\ell_1 - \alpha_1 - m + 2 \mu^{(0)}_+ + \beta_1) \bigg\}
\nonumber \\
&& \times \exp \bigg\{ - (\ell_2 - \alpha_2 - 2 \mu^{(0)}_+ + \beta_2)
\ln (\ell_2 - \alpha_2 - 2 \mu^{(0)}_+ + \beta_2) \bigg\} \ .
\label{37}
\ea
Using the same approximations to calculate $N_+$, we obtain a result
of the form~(\ref{37}) but with $\alpha_1, \alpha_2, \beta_1, \beta_2$
everywhere (including the definitions of $\tau$ and $\mu^{(0)}_+$)
replaced by zero.

We turn to the negative-parity states. For these states, the word
``even'' in Eq.~(\ref{33}) is replaced by ``odd'' and $N_+$ by
$N_-$. The calculation is completely analogous except for the
replacements $\beta_1 \to \beta_1 + 1$ and $\beta_2 \to \beta_2 -
1$. For the maximum of the integrand, that implies $2 \mu^{(0)}_- = 2
\mu^{(0)}_+ - 1$. As a consequence, the terms $2 \mu^{(0)} + \beta_1$
and $2 \mu^{(0)} - \beta_2$ have the same values for states with
positive and with negative parity. This in turn implies that the
widths $\tau$ and the terms in the exponential in
expression~(\ref{37}) have the same values for states with positive
and with negative parity. It follows that in our approximation every
weight factor for states with positive parity has the same value as
the corresponding weight factor for states with negative parity. This
result is valid beyond the Gaussian approximation used in obtaining
Eq.~(\ref{37}). Indeed, the fundamental form~(\ref{34}) depends on
$\mu$ only through the invariant combinations $\mu + \beta_1$ and $\mu
- \beta_2$.  Modifications can arise only in cases where the limits of
integration (which depend on $\alpha_1, \alpha_2, \beta_1$, and
$\beta_2$) play a role, i.e., for small values of $\ell_1, \ell_2$, or
$m$.

We have shown that in the dilute limit and for every realization of
our random-matrix model, both the first and the second moments of $H$
coincide in leading order for states with positive and for states with
negative parity. The same is true of the matrix
dimensions $N_+$ and $N_-$. Thus for every realization, our
Eqs.~(\ref{1}) to (\ref{3}) predict equal values for the ground-state
energies for both parities. How reliable is that prediction? We recall
that in the dilute limit, the average spectrum of the embedded random
two-body ensemble (EGOE(2)) is Gaussian~\cite{Mon75}. The proof given
in Ref.~\cite{Mon75} applies likewise to our model. We expect,
therefore, that in the dilute limit and to a very high degree of
approximation the spectrum of any given realization of the ensemble
has also Gaussian shape. (For a single realization, the shape of the
spectrum is defined by taking local averages over a number $n \ll N_\pm$
of neighboring levels.) That expectation rests on the plausible
assumption that our random-matrix model is ergodic, at least in the
dilute limit, and implies that for every realization, our
Eqs.~(\ref{1}) to (\ref{3}) become even better approximations as the
matrix dimension increases. We conclude that the probabilities for
ground states of positive and for negative parity are equal in the
dilute limit. That conclusion holds with the following proviso. A
preference for ground states of, say, positive parity might occur if
the local spectral fluctuation properties of the two ensembles were
locked in such a way that the positive-parity ground state fluctuates
more often towards smaller energies than does its opposite number. In
the next section we exclude that possibility. We do so by
investigating higher moments of $H$.

\section{Spectral Fluctuations}
\label{spec}

Given the coincidence of both the first and second moments of $H$ for
states of either parity in the dilute limit, we ask: does that
coincidence extend to all higher moments so that the local spectral
fluctuation properties of both ensembles are completely locked? We
approach the answer by studying higher moments of $H$.

We consider ${\rm nTrace} (H^k {\cal P}_{\pm})$ for $k$ integer and $k
\geq 3$. These traces are now shown to have the same structure as the
first and second moments of $H$: each trace is a sum of terms each of
which is the product of a monomial (or polynomial) of order $k$ in the
two-body matrix elements (the same for the projectors ${\cal P}_+$
and ${\cal P}_-$) and a weight factor that does not depend on the
random variables but may have a different value for positive and
negative parity.

We proceed as in Section~\ref{traces} but are interested only in the
general form of the result. The operator $H^k$ is a monomial of order
$k$ in the matrix elements $V^{(1)}, V^{(2)}, X^{(1)}, X^{(2)}$.  Each
matrix element carries four indices. Thus, in $H^k$ there occur $4k$
independent summations over single-particle level indices.
Non-vanishing contributions to the trace of $H^k$ arise only from
Wick-contracted terms. Each pairwise Wick contraction of a creation
and an annihilation operator in $H^k$ produces a factor of the form
$n_{1 \alpha}$, $(1 - n_{1 \alpha})$, $n_{2 \alpha}$, or $(1 - n_{2
\alpha})$, as the case may be. At the same time, two summation indices
become equal. After all Wick contractions are done, $H^k$ contains at
most $2k$ independent summations over level indices. (That number may
be smaller than $2k$ since two or more of the resulting factors $n_{1
\alpha}$, $(1 - n_{1 \alpha})$, $n_{2 \alpha}$, or $(1 - n_{2
\alpha})$ may carry the same index.) By using the identity $n^2 = n$
for the number operator, the Wick-contracted $H^k$ can be written in
such a way that the summation indices on all such factors are
different. For $k = 2$, that was done in Eq.~(\ref{12}). We consider a
single term resulting from this procedure and denote by $k_1, k_2,
k_3, k_4$ the powers of the four types of factors (in the same
sequence as listed above) in that term. The maximum power with which
all factors jointly can appear, is $2k$ so that $k_1 + k_2 + k_3 + k_4
\leq 2k$. Clearly we must also have $k_1 + k_2 \leq \ell_1$ and $k_3 +
k_4 \leq \ell_2$. We conclude that a general term in the
Wick-contracted form of $H^k$, characterized by the four integers
$k_1, k_2, k_3, k_4$ as constrained above, has the form
\ba
&& \sum_{\alpha_1, \alpha_2, \ldots, \alpha_{k_1}} \sum_{\beta_1,
\beta_2, \ldots, \beta_{k_2}} \sum_{\gamma_1, \gamma_2, \ldots,
\gamma_{k_3}} \sum_{\delta_1, \delta_2, \ldots, \delta_{k_4}}
\nonumber \\
&& \bigg\{ \prod_{r = 1}^{k_1} n_{1 \alpha_r} \prod_{s = 1}^{k_2}
(1 - n_{1 \beta_s}) \prod_{t = 1}^{k_3} n_{2 \gamma_t} \prod_{u =
1}^{k_4} (1 - n_{2 \delta_u}) \nonumber \\
&& f_{\alpha_1, \ldots, \alpha_{k_1}; \beta_1, \ldots, \beta_{k_2};
\gamma_1, \ldots, \gamma_{k_3}; \delta_1, \ldots, \ldots_{k_4}}
\bigg\} \ .
\label{38}
\ea
The sums in this expression are jointly constrained by the condition
that no two summation indices are equal. The form of the function $f$
depends upon the value of $k$. $f$ is a monomial of order $k$ in the
matrix elements $V^{1}, V^{2}, X^{1}, X^{2}$. These carry the
summation indices. The Wick-contraction of $H^k$ yields a sum of
terms of the form~(\ref{38}). For the calculation of ${\rm nTrace}
[ H^k {\cal P}_{\pm} ]$, we observe that the expression
\be
\Pi_{\pm}(k_1,k_2,k_3,k_4) = {\rm nTrace} \bigg\{ \prod_{r = 1}^{k_1}
n_{1 \alpha_r} \prod_{s = 1}^{k_2} (1 - n_{1 \beta_s}) \prod_{t =
1}^{k_3} n_{2 \gamma_t} \prod_{u = 1}^{k_4} (1 - n_{2 \delta_u})
{\cal P}_{\pm} \bigg\}
\label{39}
\ee
does not depend on the values of the indices $\alpha_1, \ldots,
\delta_{k_4}$. Therefore, the normalized traces of the projections
of the expression~(\ref{38}) are given by
\ba
&& \sum_{\alpha_1, \ldots, \alpha_{k_1}} \sum_{\beta_1, \ldots,
\beta_{k_2}} \sum_{\gamma_1, \ldots, \gamma_{k_3}} \sum_{\delta_1,
\ldots, \delta_{k_4}} f_{\alpha_1, \ldots, \alpha_{k_1}; \beta_1,
\ldots, \beta_{k_2}; \gamma_1, \ldots, \gamma_{k_3}; \delta_1,
\ldots, \ldots_{k_4}} \nonumber \\
&& \qquad \times \Pi_{\pm}(k_1,k_2,k_3,k_4) \ .
\label{40}
\ea
Expression~(\ref{40}) shows that the results derived in
Section~\ref{traces} for ${\rm nTrace} [ H {\cal P}_{\pm} ]$ and for
${\rm nTrace} [ H^2 {\cal P}_{\pm} ]$ hold for arbitrary powers $k$ of
$H$: Each trace ${\rm nTrace} [ H^k {\cal P}_{\pm} ]$ is a sum of
terms; every term in the sum is the product of two factors. The first
factor contains the random variables and is the same for the states
with positive and with negative parity. The second factor, a weight
factor, may depend on parity. We have, thus, shown that the
Hamiltonians for states with positive and with negative parity are
very highly correlated.

This is a remarkable result in its own right. Indeed, with increasing
values of $\ell_1$ and $\ell_2$ the matrix dimensions $N_+$ and $N_-$
grow approximately like $((\ell_1 + \ell_2)/m)^m$ while the number of
two-body matrix elements only grows like $(\ell_1 + \ell_2)^4$. Thus,
for $(\ell_1 + \ell_2) > m^2$, the matrix dimensions become
asymptotically very much larger than the number of independent matrix
elements. Still, in the sense of Eq.~(\ref{40}), the two Hamiltonians
remain totally correlated.

We turn to the weight factors appearing in Eq.~(\ref{40}) and show
that these are also asymptotically equal. Our statement applies up to
a maximum value of $k$ which we determine approximately. The weight
factors $\Pi_{\pm}(k_1,k_2,k_3,k_4)$ are explicitly given by
\be
\Pi_+(k_1,k_2,k_3,k_4) = \frac{1}{N_+} \sum_{m_1, m_2} \delta_{m_1 +
m_2, m} \delta_{m_2, {\rm even}} {\ell_1 - k_2 \choose m_1 - k_1}
{\ell_2 - k_4 \choose m_2 - k_3}
\label{41}
\ee
and
\be
\Pi_-(k_1,k_2,k_3,k_4) = \frac{1}{N_-} \sum_{m_1, m_2} \delta_{m_1 +
m_2, m} \delta_{m_2, {\rm odd}} {\ell_1 - k_2 \choose m_1 - k_1}
{\ell_2 - k_4 \choose m_2 - k_3} \ .
\label{42}
\ee
In the summations over $m_1, m_2$, we obviously must have $m_1 \geq
k_1$ and $m_2 \geq k_3$.  Since $m_1$ and $m_2$ are both bounded by
$m$, that condition in fact limits $k_1$ and $k_3$. It is obvious that
for large values of $k$, the two weight factors cannot always be
equal. Consider, for instance, the case $k_1 = 0$, $k_3 = m$. Then we
have $m_1 = 0$ and $m_2 = m$.  That implies $\Pi_+(k_1,k_2,k_3,k_4) =
0$, $\Pi_-(k_1,k_2,k_3,k_4) \neq 0$ if $m$ is odd and
$\Pi_-(k_1,k_2,k_3,k_4) = 0$, $\Pi_+(k_1,k_2,k_3,k_4) \neq 0$ if $m$
is even. To avoid such cases, we must have $k < m$.  Even then $\Pi_+$
and $\Pi_-$ may differ. This happens when the bounds on the summation
indices in Eqs.~(\ref{41}) and (\ref{42}) are relevant. We avoid these
cases by choosing $k \ll m$. We recall that the asymptotic regime is
characterized by the relations $1 \ll m \ll \ell_1, \ell_2$. We thus
require that $m$ is sufficiently large to accommodate the relation $k
\ll m$ and yet allows $k$ to assume values large compared to one. With
these assumptions, the arguments used above for $k = 1, 2$ show that
$\Pi_+ = \Pi_-$.

We have shown that in the asymptotic regime and for all $k$ with $k
\leq k_0$, the moments ${\rm Trace} \ ( H^k {\cal P}_{\pm})$ pairwise
have the same values for states with positive and with negative
parity. Here $k_0$ obeys $1 \ll k_0 \ll m$. That conclusion does not
depend on assuming any symmetry such as $\ell_1 = \ell_2$. We have
also shown that for $k \gg k_0$, the moments differ. As $k$ increases,
the bounds on the summations over products of binomial factors become
ever more important. As a consequence, the differences between moments
for states with positive and with negative parity increase with $k$.
That statement is relevant for the local spectral fluctuation
properties of both ensembles. Indeed, it is known~\cite{Bro81} that
such fluctuation properties depend on the very highest moments of $H$:
In the limit of infinite matrix dimension, there exists a clear
separation between the overall shape of the spectrum (defined by
averaging over an energy interval large compared to the average level
spacing $d$), and the local spectral fluctuations (defined on a scale
of order $d$). Since the moments of $H$ for states of positive and
negative parity differ for $k \gg k_0$, we conclude that the local
fluctuation properties of both ensembles are uncorrelated in the
dilute limit, even though the moments of $H$ for both parities
coincide up to $k \approx k_0$. This excludes the possibility
mentioned in Section~\ref{dil} that the local spectral fluctuation
properties of the two ensembles are locked in such a way that the
positive-parity ground state fluctuates more often towards smaller
energies than does its opposite number or vice versa and completes the
proof that in the dilute limit, ground states of either parity carry
equal probabilities.

\section{Summary and Discussion}
\label{sum}

We have shown that in the dilute limit, ground states of either parity
carry equal probabilities. That conclusion is based on the following
facts. (i) The spectra are asymptotically Gaussian, and Eqs.~(\ref{1})
to (\ref{3}) become asymptotically strictly valid. (ii) The first and
second moments of $H$ and the dimensions of the Hamiltonian matrices
become asymptotically equal for either parity so that Eqs.~(\ref{1})
to (\ref{3}) predict equal probabilities for either parity. (iii) The
local spectral fluctuation properties of the two spectra are
asymptotically uncorrelated because very high moments have different
values. That fact excludes a locking of these fluctuations.

Deviations from equal ground-state probabilities are, thus,
finite-size effects. For values of the parameters $m, \ell_1, \ell_2$
that are sufficiently small for numerical simulations, we have indeed
found such deviations. They occur whenever the dimensions $N_+$ and
$N_-$ differ. Conversely, for $N_+ = N_-$ we have not found
significant deviations from equal probabilities. The small
fluctuations found for $r_{\pm}$ in the fits to the data show that
Eqs.~(\ref{1}) and (\ref{2}) are approximately valid: They do predict
correctly which parity has the higher probability to furnish the
ground state. The values of the predicted probabilities are
semi-quantitatively correct.

Calculations using the two--body random ensemble (TBRE) reported in
Ref.~\cite{Pap06} displayed correlations between spectra carrying
different quantum numbers. One may argue that these results contradict
our present findings. This is not the case: Calculations using the
TBRE are neccessarily restricted to small matrix dimensions while our
argument for independence of spectral fluctuation properties of states
of positive and negative parity applies only in the dilute limit,
i.e. for infinite matrix dimension.

Our results may have interesting implications for the statistical
theory of nuclear reactions. There an open question is this: are
$S$-matrix elements carrying different quantum numbers like total spin
uncorrelated? That assumption is always used in the theory and is
consistent with the observed symmetry of compound-nucleus cross
sections about 90 degrees in the center-of-mass system. Still, the
assumption is not obviously valid for a realistic random-matrix model
of nuclear reactions. Normally the statistical theory of nuclear
reactions uses the Gaussian orthogonal ensemble (GOE). It would be
more realistic to use instead the TBRE.  The TBRE differs from the GOE
in that it employs a shell-model in which the two-body matrix elements
are the random variables. (For a review of the TBRE, see
Ref.~\cite{Pap07}.) But then it is the same set of random variables
that govern scattering matrix elements carrying different quantum
numbers; just as in the model considered above the same random
two-body matrix elements govern the Hamiltonians for states of
different parity. To approach the question, we observe that for
orthogonally invariant ensembles, universality holds also for elements
of the scattering matrix carrying identical quantum
numbers~\cite{Hac95}. That statement implies that correlations between
such elements depend only on local spectral fluctuation
properties. This conclusion is supported by the explicit calculation
in Ref.~\cite{Ver85} of the correlation function of a pair of
$S$-matrix elements: Aside from the strength of the coupling to the
open channels, the correlation depends solely on the value of the
local mean level density. If we assume that these statements carry
over to the TBRE, and if we further assume that in the TBRE just as in
the model studied above, the local spectral fluctuation properties of
spectra carrying different quantum numbers are uncorrelated in the
limit of large matrix dimension, we are led to the conclusion that
$S$-matrix elements carrying different quantum numbers are, likewise,
uncorrelated. The limit of a large matrix dimension is appropriate
because the resonances relevant in the statistical theory correspond
to states above the ground state.

\section*{Acknowledgments}

This work was partially supported by the U.S. Department of Energy
under contract No. DE-AC05-00OR22725 with UT-Battelle, LLC (Oak Ridge
National Laboratory), and under grant No. DE-FG02-96ER40963
(University of Tennessee).

\end{document}